\documentclass[10pt]{iopart}
\usepackage{iopams,setstack}
\usepackage[dvips]{graphicx}
\usepackage[scriptsize,bf]{caption}
\usepackage{xspace}
\usepackage{times}
\eqnobysec

\renewcommand{\br}[1]{\left(#1\right)}
\newcommand{\ud}[1]{\mathrm{d}#1}
\renewcommand{\exp}[1]{e^{#1}}
\newcommand{\sq}[1]{\left[#1\right]}

\newcommand{\abs}[1]{\left|#1\right|}

\newcommand{\eqref}[1]{\eref{#1}}

\newcommand{\dx}{\dot{x}}

\newcommand{\dq}{\dot{q}}
\newcommand{\dk}{\dot{k}}

\newcommand{\elm}{m}

\newcommand{\neqv}{\not\equiv}
\newcommand{\pb}[1]{[{#1}]_{PB}}

\newcommand{\cmf}{\textsl{CM\xspace}}
\newcommand{\dof}{degrees of freedom\xspace}

\begin{document}
\title[{\L}. Bratek \ \ \ \ \ \ \ \ Hamiltonian formulation of Relativistic Rotators ]{Can rapidity become  a gauge variable?
 Dirac Hamiltonian method and  Relativistic Rotators.  }
\author{{\L}ukasz Bratek}
\address{The H. Niewodnicza{\'n}ski Institute of Nuclear Physics,
Polish Academy of Sciences, Radzikowskego 152, PL-31342 Krak{\'o}w, Poland}
\ead{lukasz.bratek@ifj.edu.pl}

\begin{abstract}\\
The minimal Hamiltonian for a family of relativistic rotators is constructed
by a direct application of the Dirac procedure for constrained systems. The
Hamiltonian equations can be easily solved. It is found that the resulting
motion is unique and qualitatively the same for all phenomenological rotators,
only the relation between mass and spin is different. There is a critical point
in the construction when such a relation cannot be established, implying that
the number of primary constraints is greater. In that case the mass and the
spin become unrelated, separately fixed parameters, and the corresponding
Hamiltonian changes qualitatively. Furthermore, a genuine physical observable
becomes a gauge variable. This paradoxical result is consistent with the fact
already known at the Lagrangian level that the Hessian rank is lower than
expected, and the equations of motion indeterminate on $\mathbb{R}^3\times\mathbb{S}^2$.
\end{abstract}
\medskip
\hrule\medskip
\noindent
\textbf{The definitive version is available online at
\texttt{stacks.iop.org/JPhysA/45/055204}}
\hrule


\section{Historical background and motivation}

A geometric model of a spinning particle with fixed mass and spin both in a Lagrangian and
a Hamiltonian frame was proposed in \cite{bib:segal}. A solution called general with a fixed frequency
was presented. The Hamiltonian found therein becomes the basis for the quantization of the
model. Later, a short paper was written in which a family of relativistic rotators was proposed
in a Lagrangian frame \cite{bib:astar1}. The two papers converge on a final particular Lagrangian defining
the same dynamical system with separately fixed Casimir mass and spin.

Next, it was shown in \cite{bib:bratek_arXiv_1} that a single and quite arbitrary function of the time appears in the
general solution and that the fact is associated with that the Hessian rank of the Lagrangian is
only $4$, not $5$ as expected originally for that system. In the language of the Dirac formalism, this
means that the secondary constraint is trivial $0=0$ in the free motion. However, a nontrivial
secondary constraint will usually appear when an interaction term is added incapable of removing the Hessian singularity. In particular, depending on the structure of the interaction,
such a constraint could impose nonphysical limitations on the freedom in choosing initial data.

The arbitrariness in the frequency was also encountered in \cite{bib:kassandrov}; however, the authors did
not try to understand the reason for the singularity and did not pay due attention to it. They
minimally coupled the rotator with the electromagnetic field and a constraint indeed appeared.
But this is what was to be expected on account of the fact that the particular form of the
interaction was insufficient to remove the Hessian rank deficiency \cite{bib:bratek_ArXiv_EMF}. With this constraint,
the particular motion in the uniform magnetic field discussed in \cite{bib:kassandrov}
indeed was unique. But
later, with the help of a toy model with the same kind of Hessian singularity, it was illustrated
that this uniqueness was accidental and the constrained motion would be still non-unique in
the general electromagnetic field \cite{bib:bratek_JPA_toy}.

In the meantime, a paper was published \cite{bib:subir} in which a Hamiltonization scheme alternative
to that suggested in \cite{bib:segal} (and different from that we present later for all rotators) was proposed
for the particular fundamental rotator. This scheme employed additional auxiliary variables not
present in the original Lagrangian and resulted in a non-commutativity structure on which the
authors focused. The motivation for introducing these variables was the author's argumentation
that the Lagrangian in the form presented by Staruszkiewicz with its complicated structure
of time derivatives and the presence of nested square roots were not suitable for Hamiltonian
analysis due to troubles expected to arise in trying to express velocities as functions of
momenta. However, it is not clear what kind of troubles they meant precisely, since the
Dirac Hamiltonian method, as we shall see, can be applied directly, without the necessity
of introducing the additional variables. Also, the earlier Hessian deficiency result of
\cite{bib:bratek_arXiv_1}, was
probably not meant, since it was not referred to in their paper (nor the arbitrary frequency
problem). The result of \cite{bib:bratek_arXiv_1} says that even when the action integral has been expressed in
terms of only the degrees of freedom on $\mathbb{R}^3\times\mathbb{S}^2$ specific of a rotator, the corresponding
momenta could not be inverted for velocities. But this singularity is not a problem for the
Dirac procedure at all; it even predicts the fourth primary constraint not present in the case of
phenomenological rotators.

The issue of the indeterminate motion of the fundamental rotator was described
exhaustively in \cite{bib:bratek_ArXiv_EMF}. Later, we aim to elucidate this issue at the Hamiltonian level. To this
end, we carry out a Hamiltonization appropriate for all relativistic rotators by employing the
Dirac method exactly as is described in his \textit{Lectures on Quantum Mechanics}
\cite{1964Dirac}. We shall see
the important difference between the motion of fundamental and phenomenological rotators.
Moreover, the Dirac procedure will reveal, in the form of a paradox, another surprising nature
of the indeterminate motion.

\section{General remarks}

Any isolated relativistic dynamical system has ten independent integrals of motion associated
with Poincar\'{e} symmetries. The most important and meaningful functions of the integrals
are Casimir invariants of the Poincar\'{e} group, defining the mass and the spin in a covariant
manner. They are used along with other quantum numbers to identify quantum particles and
other irreducible quantum states. The Casimir invariants of the Poincar\'{e} group are of primary
importance for the quantum formalism. The Wigner irreducibility idea
\cite{1939Wigner}, which is something
pertinent to relativistic quantum systems, can also be realized at the classical level by the
requirement of the fixed mass and spin for classical relativistic systems. From this standpoint,
it is convenient and physically desirable to divide classical relativistic dynamical systems
into two classes: \textit{phenomenological} and \textit{fundamental}. Fundamental are those whose Casimir
invariants are parameters, whereas phenomenological are those whose Casimir invariants are mere integrals of motion. This important distinction between the two kinds of classical
relativistic dynamical systems was suggested by Staruszkiewicz in \cite{bib:astar1}.

Since their mass and spin are parameters, like for quantum particles, classical fundamental
dynamical systems are natural candidates for relativistic particle models. The next stage toward
the approximate description of real particles is the quantization of such models. To this end
one needs a Hamiltonian. It is difficult to find a relativistic model in the standard Hamiltonian
frame. The best suited frame for incorporating various kinds of symmetries is the Lagrangian
frame (for example, a relativistic invariant action by construction leads to relativistically
invariant equations of motion). It is also of primary importance that the mass and the spin are
defined at the Lagrangian level, since the Casimir invariants are constructed from canonical
momenta resulting from relativistic symmetries. From this standpoint, the Lagrangian level
can be considered primary, whereas the Hamiltonian level, which gives a convenient way of
dealing with the equations of motion, is secondary. However, the Hamiltonian frame is better
suited for quantization.

It is important that one correctly finds the Hamiltonian corresponding to a given
Lagrangian. One must be sure that the Lagrangian and the Hamiltonian frames are mutually
invertible. This task is challenging when constraints are present but difficult to identify.
Recall that according to Dirac the first step toward constructing a Hamiltonian is to acquire
the knowledge about all primary constraints. Primary constraints follow from the definition
of momenta. (Some of the constraints are imposed already at the stage of the Lagrangian
construction as subsidiary conditions, cf the \ref{app:example}). Sometimes, it is difficult to detect
all constraints as they can be overlooked. A good example is the fundamental relativistic
rotator. When regarded as a non-degenerate rotator, one can easily detect only the constraints
characteristic of all relativistic rotators, such as the reparametrization and projection invariance
constraints. In finding primary constraints, it is helpful to know the rank of the Hessian matrix
present in the Lagrangian equations of motion. The number of primary constraints can be
deduced from the Hessian rank. Unfortunately, the task of the determination of the Hessian
rank can be cumbersome and computationally challenging
\cite{bib:bratek_JPA_breath}.

\section{\label{sec:issue}Relativistic rotators and the issue of constraints}
According to Staruszkiewicz a relativistic rotator is a dynamical system described
in spacetime by position $x$ and a single null direction $k$ and, additionally, by two parameters, $\elm$ (mass) and $\ell$ (length) \cite{bib:astar1}. This leads to the
following most general form of action defining
a family of relativistic rotators:
 \begin{equation}\fl \label{eq:action}
S=-m\int\ud{\tau}\sqrt{\dx\dx}\sqrt{G(Q)}, \qquad \mathrm{where} \qquad
Q=\sqrt{-\ell^2\frac{\dk\dk}{(k\dx)^2}},\quad kk=0, \end{equation}
with $G$ being some positive function enumerating the rotators. Here, $\tau$ is
an arbitrary variable along the worldline.

Since $k$ is null, the first primary constraint is
$\varphi_1=kk\approx0$.\footnote{We use $\approx$ to denote a weak equation in the sense of definition given in \cite{1964Dirac,1950Dirac}.} We know it prior to
calculating momenta which read\footnote{We assume the signature ${+,-,-,-}$ for the metric tensor. Correspondingly, we use definitions of momenta with a 'minus' sign: e.g. $p=-\partial_{\dot{x}}L$.}
$$p=m\sqrt{G(Q)}\,\frac{\dx}{\sqrt{\dx\dx}}-\frac{m}{2}\frac{Q\,G'(Q)}{\sqrt{G(Q)}}\sqrt{\dx\dx}\,
\frac{k}{k\dx},
\qquad\chi=\frac{m}{2}\frac{Q\,G'(Q)}{\sqrt{G(Q)}}\sqrt{\dx\dx}\,\frac{\dk}{\dk\dk}.$$
The action is projection invariant -- the transformation $k\to k_{\lambda}=
\exp{\lambda}\, k$, with the arbitrary function $\lambda$ being defined on the worldline, is
a symmetry of the Lagrangian. Only the null direction assigned to $k$ is physically relevant. The
corresponding projection invariance constraint is $\varphi_2=\chi k$. Its presence
is the consequence of the definition of a momentum $\chi$ and that $k\dk=0$.

There is something to be cautious about. A physical state should be projection invariant
since the arbitrary factor of $k$ is effectively absent from the action integral \eqref{eq:action}. The momentum $\chi$ is not a reparametrization invariant,
it transforms as
$$\chi\quad\to\quad
\chi_{\lambda}=\exp{-\lambda}\br{\chi+\frac{m}{2}\frac{Q\,G'(Q)}{\sqrt{G(Q)}}
\sqrt{\dx\dx}\,\dot{\lambda}\, \frac{k}{\dot{k}\dot{k}}}.$$
Thus, it is not a classical observable. For
example, one could add to $\chi$ anything proportional to $k$ without altering the
physical state, and the condition $k\chi=0$ would be still satisfied (this can be
realized by adding to the Lagrangian a total derivative
$\dot{\Sigma}\,kk+2\Sigma\, k\dk$ which vanishes on the constraint surface, while
at the Hamiltonian level we have the constraint $k\chi=0$ which defines $\chi$ to
within the addition of a quantity proportional to $k$). For that reason, it is
justified to expect that momenta $\chi$ derived at the Lagrangian and at the Hamiltonian level will
not be identical but will be equal to each other only to within an additive term proportional
to $k$, and therefore some caution is required when the two levels (although equivalent) are
compared. This remark can be generalized -- the momenta defined for a dynamical system at
the Lagrangian and at the Hamiltonian level might not be identical but might be equal to each
other to within a gauge transformation.

The action integral \eqref{eq:action} is also reparametrization invariant in accordance with the
requirement of relativity. Related to this invariance is the fact that the Lagrangian is
homogeneous of the first degree in the velocities
$\dq^i\partial_{\dq^i}L\equiv{}L$. This tells us that the ordinary
Hamiltonian known from elementary mechanics vanishes identically. The corresponding
reparametrization invariance constraint is difficult to find for general $G$, but we are guaranteed
of its existence. Differentiation of the homogeneity relation gives
$\delta^i_j\partial_{\dq^i}L+\dq^i\partial_{\dq^j}\partial_{\dq^i}L=\partial_{\dq^j}L$;
hence $\dq^i\partial_{\dq^i}p_j=0$, which means that the Jacobian
$\partial_{\dq^i}p_j$ of a map from velocities to momenta
has a lower rank than the number of velocities and that there must exist at least one constraint
involving momenta, distinct from constraints $\varphi_1$ and $\varphi_2$.

Above, we have detected the three constraints characteristic of all relativistic rotators by
examining the general transformation properties of the action integral
\eqref{eq:action}. With this number
of constraints one could expect $8-3=5$ \dof
uniquely defining the physical
state of a rotator (here, $8$ is the dimensionality of the full configuration space in the action
integral). But the number of constraints can be greater, depending on the function $G$ in the
action integral, which can be inferred from the behavior of Casimir invariants. In units of $\elm$ and
$\ell$, the Casimir invariants of the Poincar\'{e} group are\footnote{We use the definition $C_J=\frac{WW}{-\frac{1}{4}m^4\ell^2}$ assuming $kk=0$ and $k\chi=0$, where  $${WW}=-
 \left|\begin{array}{ccc} pp & p\chi & pk \\ \chi p &
\chi\chi & \chi k \\ kp & k\chi & kk \end{array}\right|$$ is the square of Pauli-Luba\'{n}ski spin
pseudo-vector. In the literature, one can often encounter incorrect spelling \textit{Lubanski} instead of \textit{Luba\'{n}ski}.\\ \indent Historical note: J\'{o}zef Kazimierz Luba\'{n}ski  defined the vector known as Pauli-Luba\'{n}ski
(pseudo)-vector \cite{bib:sredniawa}. It seems that the list of inventors of the vector should be extended and it would be more appropriate to refer to it as
\textsl{Mathisson-Pauli-Luba\'{n}ski spin pseudovector}:
\textit{ Sometime in the 1960s, Weyssenhoff told his then Ph.D. student Andrzej Bia{\l}as
that it was Mathisson who had explained to Luba\'{n}ski how to construct, from
the spin bivector $s^{\mu\nu}$, the object that is now known as the Pauli-Luba\'{n}ski
vector \cite{bib:trautman}.}
} \begin{equation}
\label{eq:CasInvMom}C_M := \frac{pp}{m^2}>0,\qquad C_J :=
\frac{\chi\chi\,\br{pk}^2}{-\frac{1}{4}m^4\ell^2}>0.\end{equation} Using the
definitions of momenta, the invariants can be expressed in terms of $G(Q)$:
$C_M=G(Q)-QG'(Q)$  and $C_J=G'(Q)^2$, or alternatively as
\begin{equation}\label{eq:InvDepOnQ} {C_M\pm Q\sqrt{C_J}}=G(Q),\qquad
\pm\sqrt{C_J}=G'(Q).\end{equation} Function $G>0$ cannot be arbitrary, it should satisfy the following requirement: $$G(Q)>QG'(Q)\not\equiv0.$$ The condition  $G(Q)>QG'(Q)$ is a
consequence of positivity of the square of mass, $C_M>0$, whereas   the condition
$G'(Q)\not\equiv0$ indicates we are excluding the case of structureless point
particle. A variation in $Q$ will cause the corresponding variations to occur in
the Casimir invariants: $\delta{C_M}=-Q\,G''(Q)\,\delta{Q}$ and
$\delta{C_J}=2\,G'(Q)\,G''(Q)\,\delta{Q}$. Hence, if $G''(Q)\neqv0$, the Casimir
invariants are functions of each other, and there is a $G$-dependent function
$F_G\br{C_M,C_J}=0$ involving momenta and positions through three scalars: $pp$,
$pk$ and $\chi\chi$. This gives us the third reparametrization invariance constraint
$$\varphi_3=F_G(C_M,C_J)\approx0,\qquad G''(Q)\not\equiv0$$  existence of which we
already know. Thus, when $G''(Q)\not\equiv0$, we indeed have three constraints for
eight \dof in agreement with the expectation that a rotator is a dynamical system
with five (physical) \dof.

When $G''(Q)\equiv0$, the situation is qualitatively different and somewhat degenerate. In that
case, the constraint $F_G(C_M,C_J)=0$ is no longer valid, since the necessary condition for the
invertibility of $C_J=G'(Q)^2$ is broken. Instead, we must use $C_M=G(0)>0$ and $C_J=G'(0)^2$ separately; only then relations
\eqref{eq:InvDepOnQ} can be satisfied. Then, both the Casimir invariants are
identically fixed, independently of the initial conditions, in which case the Casimir mass and
spin are the parameters, not merely the integrals of motion. Without loss of generality, we can
then put both the $C_M$ and $C_J$ equal to $1$, since fixed scales can be absorbed by dimensional
constants $\elm$ and $\ell$ which have yet been
unspecified. This gives us $G(Q)=1\pm{}Q$. This way
we have arrived at the Lagrangians of two degenerate rotators
\begin{equation}\label{eq:action_fr}S=-m\int{\ud{\tau}\sqrt{\dot{x}\dot{x}}
\sqrt{1\pm\sqrt{-\ell^2\frac{\dot{k}\dot{k}}{\br{k\dot{x}}^2}}}}.\end{equation}
Originally, the Lagrangians were found by imposing the condition of fixed mass and spin \cite{bib:segal,bib:astar1}, and later, by requiring that the Hessian
for the rotator degrees of freedom on $\mathbb{R}^3\times\mathbb{S}^2$ should be
singular \cite{bib:bratek_arXiv_1,bib:bratek_ArXiv_EMF}. Now, we have obtained them as the critical case when the
reparametrization constraint of phenomenological rotators is no longer valid in the form of $\varphi_3$. One can say, there is a transition in the number of constraints when function $G(Q)$ becomes
linear, in which case the reparametrization invariance constraint splits into two independent
constraints. This phenomenon is consistent with the result of \cite{bib:bratek_arXiv_1,bib:bratek_ArXiv_EMF} implying that there must
be $8-4=4$ primary constraints when $G''(Q)\equiv0$ (and
$G'(Q)\not\equiv0$) on account of the
fact that then the Hessian rank is $4$. A similar argument shows that we have $8-5=3$ primary constraints when $G''(Q)\not\equiv0$, since then the Hessian rank is $5$ (the connection between the
Hessian rank and the number of primary constraints is illustrated in a simple model in the
\ref{app:example}).

\section{The Hamiltonian for phenomenological rotators}
All three primary constraints found for phenomenological rotators are regular and independent
on the constraint surface. This can be verified by ascertaining if the rank of the Jacobian of a map from the phase space coordinates to constraints regarded as new coordinates is the same
as the number of constraints. Instead, one can check that a $3\times3$ matrix with elements \begin{equation}\label{eq:Jacob}\mathcal{J}_{mn}=\eta^{\mu\nu}\br{ \alpha \frac{\partial{{\varphi}}_m}{\partial{}p^{\mu}}\frac{\partial{{\varphi}}_n}{\partial{} p^{\nu}}+ \beta\frac{\partial{{\varphi}}_m}{\partial{}k^{\mu}}\frac{\partial{{ \varphi}}_n}{\partial{}k^{\nu}}+ \gamma\frac{\partial{{\varphi}}_m}{\partial{}\chi^{\mu}}\frac{\partial{{\varphi}}_n}{ \partial{}\chi^{\nu}}}\end{equation} is non-singular on the constraint surface ($\alpha,\beta,\gamma$ are arbitrary functions introducing appropriate units). Its determinant is equal to $-16\,\beta^3\,\chi\chi\,C_J^2\,\br{F_{G,C_J}}^2$ on the constraint surface and cannot be identically zero for the nonzero spin (if $F_{G,C_J}\equiv0$ then $G''\equiv0$, and we are excluding that case).  This regularity result remains unaltered if the ${{\varphi}}_m$'s are multiplied by any
functions nonzero on the constraint surface. From a calculation carried out in \cite{bib:bratek_ArXiv_EMF} it follows
that the Hessian rank is $5$ when $G''(Q)\ne0$. Since the number of velocities is $8$, we can be
sure to have found all independent primary constraints. Furthermore, $\pb{\varphi_1,\varphi_2}=2\varphi_1\approx0$, $\pb{\varphi_2,\varphi_3}=0$ and $$\pb{\varphi_3,\varphi_1}=\frac{16\br{pk}^2}{m^4\ell^2}F_{G,C_J}\,\varphi_2\approx0.$$ This means that all primary constraints are of the first class. There are no secondary constraints.
Thus, the total Hamiltonian, in accordance with the Dirac procedure \cite{1964Dirac,1950Dirac}, is a linear combination of the primary constraints $$H_T=u_1\, kk+ u_2\, k\chi+u_3\,F_G(C_M,C_J),\qquad G''(Q)\ne0$$ with arbitrary functions $u_m$, where $C_M$ and $C_J$ must be expressed in terms of momenta using \eqref{eq:CasInvMom}. Hamiltonian $H_T$ is also the full Hamiltonian describing phenomenological rotators.

In what follows we shall examine the dynamics resulting from the above Hamiltonian. Let
us consider the trajectory perceived in the \cmf frame. By definition, the trajectory is a projection
of the worldline onto the subspace orthogonal to $p$. By $\bot$ we denote a projection operation: $y$ $\to$ $y_{\bot}=y-\frac{yp}{pp}p$ for any vector $y$. From the Hamiltonian equations, $\dot{y}\equiv\mathrm{d}_{\tau}y=\pb{y,H_T}$; hence, $\dot{y}$ must be a linear combination of $p$, $k$ and  $\chi$.    Vectors $\dot{x}_{\bot}\equiv\br{\mathrm{d}_{\tau}{x}}_{\bot}$, $\ddot{x}_{\bot}\equiv\br{\mathrm{d}_{\tau}{\br{\dot{x}_{\bot}}}}_{\bot}$ and
$\dddot{x}_{\bot}\equiv\br{\mathrm{d}_{\tau}\br{\ddot{x}_{\bot}}}_{\bot}$, being $p$-orthogonal linear combinations of $k$, $\chi$ and $p$, must be coplanar. Thus, the torsion scalar vanishes for this trajectory.
The radius of curvature of the trajectory, $\rho$, is constant and fixed by the initial conditions
$$\begin{small}\rho={\frac{\br{\dot{x}_{\bot}\dot{x}_{\bot}}^{3/2}}{
\sqrt{-\left|\begin{array}{cc}\dot{x}_{\bot}\dot{x}_{\bot}&\dot{x}_{\bot}\ddot{x}_{\bot}\\ \ddot{x}_{\bot}\dot{x}_{\bot}&\ddot{x}_{\bot}\ddot{x}_{\bot}\end{array}\right|}}}
=\frac{\ell}{2}\frac{\sqrt{C_J}}{C_M},\end{small}$$ where
$\dot{x}_{\bot}=2\frac{c_3}{\elm}C_JF_{G,C_J}\frac{\elm}{kp}\br{k-\frac{kp}{pp}p}$ and $\ddot{x}_{\bot}=\frac{\dot{c}_3}{c_3}\dot{x}_{\bot}-\frac{16}{\ell C_J}\br{\frac{c_3}{\elm}C_JF_{G,C_J}}^2\frac{kp}{\ell\elm^2}\br{\chi-\frac{p\chi}{kp}k}$
on
account of the Hamiltonian equations. The trajectory is thus circular. The hyperbolic angle $\psi$ between $p$ and $\dx$ (called rapidity)
describing the rotation speed with respect to the  \cmf frame is (we may assume $pk>0$)
\begin{equation}\fl\label{eq:rapidity_def}\tanh{\psi}=\frac{{pk}\,{p\dx}-
{pp}\,{k\dx}}{{pk}\,{p\dx}}=-\sqrt{pp}\frac{n\dx}{p\dx},\qquad
n\equiv\frac{pp\,k-pk\,p}{pk\sqrt{pp}}.\end{equation} The Hamiltonian equations of
motion can be used to prove the identity
\begin{equation}\label{eq:rapidity}\tanh{\psi}=\rho\,\omega,\qquad
\omega\approx\frac{2}{\ell}\frac{C_M\sqrt{C_J} F_{G,C_J}}{{C_M
F_{G,C_M}+C_JF_{G,C_J}}},\end{equation}
which holds on the constraint surface, where we have introduced the angular velocity $\omega$.
Being
independent of the arbitrary functions $u_m$, $\psi$ is a genuine gauge-invariant and Lorenz-invariant
quantity with a definite meaning. In addition, $\psi$ remains constant during motion and depends
only on the initial conditions, similarly as any function of the Casimir mass and spin, since $\pb{C_M,H_T}\equiv0$ and  $\pb{C_J,H_T}\approx0$.
That $\psi$ is a constant and gauge independent was to be
expected; however, we stress this fact in view of the findings of the following section. There are
two conserved vectors, the momentum $p$ (since $\pb{p,H_T}=0$) and the spin vector $p\wedge k\wedge\chi$,
since $$\pb{k\wedge\chi,H_T}=k\wedge\pb{\chi,H_T}+\pb{k,H_T}\wedge
\chi=u_3\,\frac{8\, pk\, \chi\chi}{m^4\ell^2}\,F_{G,C_J}\,k\wedge p.$$ It is physically permissible to fix the three arbitrary functions $u_m$ so as
$$p\dot{x}=\sqrt{pp},\qquad pk=\sqrt{pp},\qquad p\chi=0.$$ With these supplementary conditions, the physical state will be unaltered and we are left only
with five \dof
as required for a rotator. The first gauge condition sets the arbitrary $\tau$
variable to be the proper time in the \cmf frame. The second one sets the arbitrary scale of $k$. The third one is also admissible since the constraint
$\varphi_2=k\chi\approx0$ defines $\chi$ to within an
additive term proportional to
$k$. Recall that this gauge is consistent with the Lagrangian frame
in which
$\chi\propto \dk$, that is $k\chi=0$ -- had the scale of $k$ be varied, the
momentum $\chi$ would
have been changed by an additive term proportional to $k$. With the above gauge conditions,
the description of a physical state is naturally adapted to the \cmf frame, and is still
relativistically covariant. These conditions now must be made consistent with the equations
of motion. Irrespective of any gauge, we always have $\pb{{p\dot{x}},H_T}\approx0$ and $p\dot{x}\equiv\sqrt{pp}$
gives $u_3$, whereas we will have $\pb{pk,H_T}\approx0$ and
$\pb{p\chi,H_T}\approx0$ holding only when $$
u_1=-\frac{1}{2}\frac{\br{p\chi}^2-pp\,\chi\chi}{pk\sqrt{-\chi\chi\,pp}}\,\omega,
\quad u_2=\frac{p\chi}{\sqrt{-\chi\chi\,pp}}\,\omega, \quad
u_3=\frac{m^4\ell^2}{8\,pk\sqrt{-\chi\chi\,pp}}\,\frac{\omega}{F_{G,C_J}}.$$ In order to verify that the denominator in the definition of $\omega$
\eqref{eq:rapidity} is nonzero (and so, the above $u_m$'s finite), consider a
function $u(x,y)=x F_{,x}+yF_{,y}$, and suppose that the constraint $F(x,y)=0$ can
be locally solved for $x$, that is, $x=f(y)$ and $F(f(y),y)\equiv0$. In that case,
there is a nonzero function $s(x,y)$ such that $F(x,y)=\br{x-f(y)}s(x,y)$. Hence,
$u(f(y),y)=s(f(y),y)\br{f(y)-y f'(y)}$. Suppose for contradiction that $u(x,y)=0$
on the constraint surface. Then $f(y)-y f'(y)=0$, which implies that the Casimir
invariants are proportional, $C_J=\alpha\, C_M$, $\alpha>0$. Then, relations
\eqref{eq:InvDepOnQ} imply $G'^2-\alpha\, G+\alpha\, Q\, G'=0$ and, on
differentiation, $G''\br{2G'+\alpha\, Q}=0$. This gives us either $G''=0$ or
$G=-\case{\alpha}{4}Q^2<0$. Both possibilities are in conflict with our assumptions
concerning $G$. This way we have come to the conclusion that the above gauge is
always feasible. In this gauge, the Hamiltonian equations of motion no longer
contain arbitrary functions and read \begin{eqnarray}
\label{eq:HamEqPhen}\fl\dot{x}=\frac{p}{\sqrt{pp}}+\varrho\,\omega
\,{\frac{pp\,k-pk\,p}{pk\sqrt{pp}}},\qquad \dot{p}=0,\qquad
\dot{k}=\omega\,\frac{p\chi\,k-pk\,\chi}{\sqrt{-\chi\chi\, pp}}, \nonumber\\ \fl
\frac{\dot{\chi}}{\sqrt{-\chi\chi}}=\omega\br{\frac{pp\,k-pk\,p}{pk\sqrt{pp}}+
\frac{p\chi}{\sqrt{-\chi\chi\,pp}}\cdot\frac{p\chi\,k-pk\,\chi}{pk\sqrt{-\chi\chi}}}.
\end{eqnarray} Using the \cmf gauge explicitly, the equations for the variable
quantities reduce to $$\dx=\frac{P}{\sqrt{PP}}+R\,\Omega\, n,\quad \dot{n}=-\Omega
\frac{\chi}{\sqrt{-\chi\chi}}, \quad
\frac{\dot{\chi}}{\sqrt{-\chi\chi}}=\Omega\,n,\qquad k={\frac{P}{\sqrt{PP}}+n}.$$
Here, $P$, $\Omega$ and $R$ denote the constant values of the momentum $p$ and of the
scalars $\omega$ and $\rho$, respectively, which we have already demonstrated to be constants of motion. It is thus a simple
matter to solve these equations, and the solution reads $$x(t)=\frac{P}{\sqrt{PP}}\,t+R\,\Omega\int n(t)\,\ud{t}, \qquad
\ddot{n}(t)+\Omega^2\,n(t)=0, \quad Pn=0, \quad nn=-1.$$ This is the composition of two motions -- an inertial motion of the \cmf frame and a uniform
rotation with a constant frequency $\Omega$ around the circle of radius $R$ in the \cmf frame in a
spacelike plane perpendicular to both the spacelike (Pauli-Luba\'{n}ski) spin
pseudovector and the time-like momentum vector. It is interesting to see that all phenomenological rotators move in
the same qualitative manner. Only the relation between constants $\Omega$, $R$ and Casimir mass or spin is different.

Finally, we give the Hamiltonian $H_{CM}$ defining the family of phenomenological rotators
in the \cmf gauge that leads to the Hamiltonian equations of motion
\eqref{eq:HamEqPhen} \begin{equation}\fl\label{eq:hamilt_CM_phenom}
H_{CM}=\frac{m}{2}\frac{f'(C_J)\sqrt{f(C_J)}}{f(C_J)-C_Jf'(C_J)}\sq{\frac{C_M-f(C_J)}{f'(C_J)}+4\,
\frac{\br{p\chi}^2-pp\,\chi\chi}{m^4\ell^2}\,kk-8\,\frac{pk\,p\chi}{m^4\ell^2}\,k
\chi},\end{equation} where $C_M$ and $C_J$ must be expressed in terms of momenta
using \eqref{eq:CasInvMom}. Function $f$  establishes
the dependence of the Casimir mass on the Casimir spin for a given rotator in the form $C_M=f(C_J)$, $C_J>0$,
and is related to function $G$ in the action \eqref{eq:action} in a complicated
way, $C_J=G'^2(Q)$, where $Q^2=4\,C_J\,f'^2(C_J)$ (this can be shown by
differentiating $C_M$ and $C_J$ expressed in terms of $G$). In this notation
$\tanh{\psi}=\frac{C_Jf'(C_J)}{C_J f'(C_J)-f(C_J)}$,
$\omega=\frac{\tanh{\psi}}{\rho}$ and $\rho=\frac{\ell}{2}\frac{\sqrt{C_J}}{f(C_J)}$.
The condition $|\tanh{\psi}|<1$ imposes a limitation on admissible $f$, related to
that we have already had for $G$: $C_M>0$ $\Rightarrow$ $G(Q)-QG'(Q)>0$
$\Rightarrow$ $\abs{\tanh\psi}=|\frac{QG'(Q)}{2G(Q)-QG'(Q)}|<1$.

\section{The Hamiltonian for fundamental rotators}

The complete set of primary constraints we have found for fundamental rotators are
$${\widetilde{\varphi}}_1=kk,\qquad {\widetilde{\varphi}}_2=\chi k,\qquad
{\widetilde{\varphi}}_3=C_J-1, \qquad {\widetilde{\varphi}}_4=C_M-1,$$ or possibly their linear combinations, which is equivalent. Similar to the previous section, we
check the correctness of the constraints. A determinant of a $4\times4$-dimensional
matrix with
elements defined by analogy with \eqref{eq:Jacob} equals to
$\frac{16\,m^2\ell^2}{(pk)^2}\alpha\beta^3$ on the constraint surface and
cannot be zero, showing that the rank of the associated Jacobian matrix is equal to the number
of constraints. The four constraints are therefore regular and independent. All the primary
constraints are the first class constraints: $\pb{{\widetilde{\varphi}}_m,{
\widetilde{\varphi}}_{m'}}\approx0$. We are now prepared to write down
the total first class Hamiltonian, which is a linear combination of the primary constraints. It is
essentially different from that previously found for phenomenological rotators. It reads $$\widetilde{H}_T=\widetilde{u}_1\, kk +\widetilde{u}_2\,
k\chi+\widetilde{u}_3\,\br{C_J-1}+
\frac{1-\widetilde{u}_4}{\widetilde{u}_4}\,\widetilde{u}_3\,\br{C_M-1}.$$ For the later convenience, we have redefined the arbitrary coefficient in the last term, which is
quite permissible on account of the fact that the functions $\widetilde{u}_m$ are
arbitrary, and $\widetilde{\varphi}_3$ and $\widetilde{\varphi}_4$ dimensionless. Owing to the fact that we have managed to detect all primary constraints, it took
only several lines to find the above Hamiltonian in the Dirac formalism. In \cite{bib:subir}, a Hamiltonian
equivalent to that we just found was arrived at through conceptually quite a different and more
intricate way, employing the additional auxiliary degrees of freedom not present in the original
Lagrangian \eqref{eq:action_fr}.

The Poisson brackets $\pb{\widetilde{H}_T,{\widetilde{\varphi}}_m}$ all vanish on
the constraint surface: $\pb{\widetilde{H}_T,{\widetilde{\varphi}}_m}\approx0$ with
the help of primary constraints. There are no further constraints -- the Hamiltonian equations
are already consistent. Functions $\widetilde{u}_m$ are completely arbitrary. All the functions have the nature
of gauge variables in the Dirac formalism, since all the primary constraints are of the first
class. Thus, there will be four arbitrary functions in the general solution for eight original
\dof. This result is unusual as it means that only four degrees of freedom are
physical, not five as expected for a rotator. The result is also consistent with the observation
made in
\cite{bib:bratek_arXiv_1,bib:bratek_ArXiv_EMF} that the Lagrangian equations of motion on
$\mathbb{R}^3\times\mathbb{S}^2$ are linearly dependent and that the general solution expressed in terms of only the coordinates on
$\mathbb{R}^3\times\mathbb{S}^2$ contains a
single arbitrary function, which is tantamount to the observation of the absence of (nontrivial)
secondary constraints we have just made at the Hamiltonian level. Moreover, with the help of
the Hamiltonian equations of motion corresponding to $\widetilde{H}_T$:
$$\dx^{}=\frac{2}{m}\,\widetilde{u}_3\,\br{n+\frac{1}{\widetilde{u}_4}\frac{p}{m}},
\qquad\dot{p}=0$$
$$\dk=\widetilde{u}_2\,k-\widetilde{u}_3\,\frac{8\br{pk}^2}{m^4\ell^2}\,\chi,\qquad
\dot{\chi}=-2\,\widetilde{u}_1\, k-\widetilde{u}_2\,\chi-\widetilde{u}_3\frac{2}{pk}\,p,$$
we come to the conclusion that the rapidity calculated according to definition
\eqref{eq:rapidity_def} is directly related to the gauge variable
$\widetilde{u}_4$, namely $\tanh{\psi}$ is numerically equal to $\widetilde{u}_4$
on the constraint surface (!): $$\tanh{\psi}\approx\widetilde{u}_4.$$
Alternatively, in
place of $\widetilde{u}_4$ we can use the angular velocity $\widetilde{\omega}$ of
rotation perceived in the \cmf frame and corresponding to $\psi$ according to the
formula $\frac{\ell}{2}\widetilde{\omega}\equiv\widetilde{u}_4\approx\tanh{\psi}$.

The rapidity is a quantity restricted to the constraint surface. It satisfies the gauge
invariance condition -- its Dirac bracket weakly vanishes with all first class constraints
(here, Poisson brackets are equivalent to Dirac brackets on account of the fact that there
is no second class constraints), that is, $\pb{\,\tanh{\psi},\widetilde{\varphi}_m}=0$ on the constraint surface (this
can be verified with the help of definition \eqref{eq:rapidity_def} and the above Hamiltonian equations
of motion used for calculating scalar products
$p\dot{x}$ and $k\dot{x}$, whilst taking care that the
constraints must not be used before working out the Poisson brackets). Thus, the rapidity
satisfies the requirements for being a classical observable in the sense of the definition
given in Henneaux and Teitelboim's handbook on \textit{Quantization of Gauge Systems}
\cite{bib:QGS}. A
physical state should not depend on gauge variables. Surely,
$\psi$ defines a physical state
and simultaneously, as we have seen, on the constraint surface it is numerically equal to
a gauge variable. This way we have come across the apparent paradox that a genuine
physical quantity $\psi$ turns out to be a genuine gauge variable! It should be stressed that 
$\psi$ is reparametrization invariant and projection invariant on the constraint surface; therefore,
the arbitrariness of
$\psi$ has nothing to do with the arbitrariness in choosing the time variable
or the scale of $k$ and is the characteristic of the fundamental rotators only and originates from
the Hessian rank deficiency discussed earlier.

Having said this, we could end this section. However, we shall find the motion of the
system described by $\widetilde{H}_T$ in order to see its correspondence with the results of \cite{bib:bratek_arXiv_1,bib:bratek_ArXiv_EMF} arrived
at the Lagrangian level. Since $\pb{p,\widetilde{H}_T}=0$ and
$pk\,\pb{k\wedge\chi,\widetilde{H}_T}=\widetilde{u}_3\,k\wedge p$, there are two
conserved vectors, the momentum $p$ and the spin vector $p\wedge k\wedge\chi$. We can make use of the
apparent symmetries of the model to set three arbitrary functions from among $\widetilde{u}_m$. Similar to
phenomenological rotators, we impose three admissible gauge conditions naturally adapted to
the \cmf frame (of which choice has been already justified in the previous section): the proper
time condition $\dx{}p\equiv{}\elm$, the projection condition $kp\equiv{}\elm$ and the
orthogonality condition $p\chi=0$. This requires the gauge variables to be set as
follows:
$$\widetilde{u}_1=-\frac{m^3}{2(pk)^2}\br{1+\frac{4(pk)^2(p\chi)^2}{m^6\ell^2}}
\widetilde{u}_4,\quad
\widetilde{u}_2=\frac{4\,pk\,p\chi}{m^3\ell^2}\,\widetilde{u}_4,\quad
\widetilde{u}_3=\frac{m}{2}\,\widetilde{u}_4. $$ In the \cmf gauge the  Hamiltonian
of fundamental rotators reads
\begin{eqnarray}\fl\label{eq:hamilt_CM_fund}\widetilde{H}_{CM}=\frac{m}{2}\sq{\frac{pp}{m^2}-1}
+\dots \nonumber\\ \fl\phantom{\widetilde{H}_{CM}}\dots
-\frac{m\ell}{4}\,\widetilde{\omega}
\sq{\br{\frac{pp}{m^2}+\frac{4(pk)^2\chi\chi}{m^4\ell^2}}+\frac{4\br{(p\chi)^2-pp\,
\chi\chi}}{m^4\ell^2}\,kk-\frac{8\,pk\,p\chi}{m^4\ell^2}\,k\chi },\end{eqnarray} where
we have expressed the function $\widetilde{u}_4$ in terms of the arbitrary function
$\widetilde{\omega}$ (as defined earlier). Recall, that the constraints assumed for
that Hamiltonian are $kk=0$, $k\chi=0$, $pp=\elm^2$ and
$(pk)^2\chi\chi=\case14{}\elm^4\ell^2$.

Any function of six independent phase-variable scalars
$kk,pk,k\chi,pp,p\chi,\chi\chi$ has its Poisson bracket with the Hamiltonian
$\widetilde{H}_{CM}$ already vanishing  on the constraint surface. The
persistent arbitrariness present in the Hamiltonian through $\widetilde{\omega}$ cannot be therefore removed by
using an argumentation from within the framework of the model. The Hamiltonian equations
of motion corresponding to $\widetilde{H}_{CM}$ are
$$\dot{x}=\frac{p}{m}+\frac{\ell}{2}\,\widetilde{\omega}
\br{\frac{m}{pk}k-\frac{p}{m}},\qquad\dot{p}=0,$$ $$\dot{k}=\widetilde{\omega}
\frac{2\,pk\br{p\chi\,k-pk\,\chi}}{m^3\ell}, \qquad
\dot{\chi}=\frac{m^2\ell}{2\,pk}\widetilde{\omega}
\br{\frac{m}{pk}k-\frac{p}{m}+\frac{4\,pk\,p\chi}{m^5\ell^2}\br{p\chi\,k-pk\,\chi}}.
$$ By using the \cmf gauge explicitly, the above equations reduce to
$$\dot{x}(t)=\frac{P}{m}+\frac{\ell}{2}\,\widetilde{\omega}(t) \,
n(t),\qquad\dot{n}(t)=-\frac{2}{m\ell}\,\widetilde{\omega}(t)\,\chi(t), \qquad
\dot{\chi}(t)=\frac{m\ell}{2}\,\widetilde{\omega}(t)\, n(t), $$ $$ nn=-1,\quad
nP=0, \quad P=const.,\quad PP=m^2,$$ where we have shown explicitly the dependence on $t$, where $t$ is the proper time measured in the \cmf frame. These equations can be easily integrated as
$$x(t)=x(0)+\frac{P}{m}\,t+\frac{\ell}{2}\int\widetilde{\omega}(t)\,n(t)\ud{t},\qquad
\br{\frac{1}{\widetilde{\omega}(t)}\frac{\ud}{\ud{t}}}^2n(t)=-n(t).$$ It is now
clear why $\widetilde{\omega}(t)$ has the interpretation of the frequency of
rotation on the unit sphere (of null directions) perceived in the \cmf frame. These
are the same solutions with the arbitrary frequency  we arrived at in the Lagrangian
frame in \cite{bib:bratek_arXiv_1,bib:bratek_ArXiv_EMF} (shown explicitly
therein). This shows that we have found the correct minimal Hamiltonian
corresponding to  the Lagrangian \eqref{eq:action_fr}.

Finally, we stress again that the frequency $\widetilde{\omega}(t)$ in the above solution is a gauge variable -- a
completely arbitrary function of the proper time in the \cmf frame, whereas there was no such
arbitrariness in the motion of phenomenological rotators for which the analogous frequency
was a constant of motion fixed by the initial data (and, of course, gauge independent).

\section{Summary and conclusions}
According to Dirac, the physical state of a system should be independent of gauge variables,
which can be arbitrary functions. This statement can be used as an argument for deciding
whether a geometric particle model can be considered physical. We applied this idea to the
family of relativistic rotators defined in
\cite{bib:astar1}. To this end, we constructed minimal Hamiltonians
for such systems using the Hamiltonian method for constrained systems suggested by Dirac
\cite{1964Dirac,1950Dirac} and solved the resulting equations of motion in the gauge adapted to the center of
momentum frame. It turned out that there are in fact two distinct kinds of rotators described
by qualitatively different Hamiltonians, namely a continuous family of phenomenological
rotators with unique and qualitatively the same motion, differing from each other in the 
spin-mass relation only, and a two-element family of fundamental rotators with separately fixed
mass and spin and indeterminate motion. In the gauge, adapted to the center of momentum
frame, the Hamiltonian for phenomenological rotators is unique, whereas the Hamiltonian for
fundamental rotators is not unique and still contains a single gauge variable.

To see clearly the nature of the singularity in the motion of the fundamental rotator, we
chose a classical observable being what particle physicists would call the rapidity with respect
to the center of momentum frame, constructed for rotators as the hyperbolic angle between the
$4$-momentum and the $4$-velocity. It provides a measure of the frequency of rotation perceived
in the center of momentum frame. Beyond all question, this observable is a genuine physical
quantity and as such should be independent of gauge variables. At the Hamiltonian level, this
expectation is confirmed for phenomenological rotators -- the rapidity is independent of gauge
variables present in the Hamiltonian, and the Cauchy problem for the Hamiltonian equations
of motion has a unique solution in the \cmf gauge.
Surprisingly, this is quite different from the
situation with the fundamental rotator \cite{bib:astar1} or, equivalently, the geometric model of the arbitrary
spin massive particle \cite{bib:segal}.
At the Hamiltonian level, we confirmed for the fundamental rotator
the result we obtained at the Lagrangian level in
\cite{bib:bratek_arXiv_1,bib:bratek_ArXiv_EMF},  that the rapidity remains completely
indeterminate and can be an arbitrary function of the time -- the secondary or Hessian constraint
present owing to the Hessian rank deficiency is absent (it is
trivial $0=0$) in the free motion. Furthermore, according to the Dirac formalism, the rapidity of the fundamental rotator (or
the associated frequency of rotation) should be regarded as a genuine gauge variable. Here, a
paradox comes about -- we have a physical degree of freedom that simultaneously is a gauge
variable. Thus, we have another way, complementary to that outlined in \cite{bib:bratek_arXiv_1} and
\cite{bib:bratek_ArXiv_EMF}, of seeing
that the fundamental rotator is defective as a dynamical system. For physical reasons, it is not
suitable for quantization, despite the fact that the minimal Hamiltonian is known (cf.\eqref{eq:hamilt_CM_fund})
and the quantization procedure could in principle be applied. Although there is a possibility
of setting the frequency at the level of the Hamiltonian (cf.
$\widetilde{\omega}$ in \eqref{eq:hamilt_CM_fund}), which is permissible
for gauge variables, this should not be done on account of the physical interpretation of
$\widetilde{\omega}$ (even when the frequency has been fixed 'by hands' in the Hamiltonian, the motion would
be unstable on account of the fact that the null space of the Hessian on the rotator manifold $\mathbb{R}^3\times\mathbb{S}^2$ is nontrivial
\cite{bib:bratek_ArXiv_EMF,bib:bratek_JPA_toy}, whereas a physical state
should be stable).

There is also another importance of our paper. It can be regarded as an illustration that
relativistic dynamical systems whose Casimir mass and spin are fixed parameters can have
at least one primary constraint more than their counterparts, with Casimir invariants being
ordinary constants of motion. This observation may be helpful in the context of finding a
well-behaved fundamental dynamical system (if it exists).

\appendix

\begin{small}
\section{\label{app:example} An example of a Lagrangian on the unit sphere illustrating the origin of constraints $\varphi_1$ and $\varphi_2$ }

In this example, we illustrate the mechanism of how constraints $\varphi_1$ and
$\varphi_2$ we encountered
for rotators come about in a simpler projection invariant model involving a null vector in
Minkowski space. To this end, we consider a particle model described by the Lagrangian $$L=-\frac{1}{2}\frac{\dot{q}\dot{q}}{\br{wq}^2}$$ with
$w$ being a constant time-like vector, and assume that positions $q$ are
constrained to a light cone $qq=0$, then always $wq\ne0$ and $\dot{q}$ is space-like: $\dot{q}\dot{q}<0$. For simplicity, we
deliberately break the reparametrization invariance. The Lagrangian is projection invariant;
the transformation $q\to \lambda q$ is a symmetry when $qq=0$. We see that there will be only two
important degrees of freedom similar to a particle constrained to a unit sphere. To find the
equations of motion in a covariant notation without introducing the internal coordinates on the
sphere, we apply the Lagrange multiplier method and add to the Lagrangian the term $\frac{1}{2}\Lambda\,
qq$ vanishing on the light-cone, with a function $\Lambda$ yet to be determined.
This gives us the extended Lagrangian and the momentum  $p$ conjugate to $q$
$$L=-\frac{1}{2}\frac{\dot{q}\dot{q}}{\br{wq}^2}+\frac{1}{2}\Lambda\, qq,\qquad
p=-\frac{\partial L}{\partial \dot{q}}=\frac{\dot{q}}{\br{wq}^2}$$ and the Lagrangian equations of motion $$qq=0,\quad
\dot{p}=-\frac{\dot{q}\dot{q}}{\br{wq}^3}w-\Lambda q,\quad \Rightarrow \quad
\frac{w\dot{p}}{qw}+\frac{\dot{q}\dot{q}}{\br{wq}^4}ww=-\Lambda.$$ The equations may
be rewritten  as
$$qq=0,\qquad\ddot{q}_{\nu}\br{\delta^{\nu}_{\phantom{\nu}\mu}-\frac{w^{\nu}q_{\mu}}{wq}-
\frac{q^{\nu}w_{\mu}}{wq}+w^2\frac{q^{\nu}q_{\mu}}{\br{wq}^2}}=2\frac{w\dot{q}}{wq}
\br{\dot{q}_{\mu}-\frac{w\dot{q}}{wq}q_{\mu}}$$ with the help of
$\dot{q}\dot{q}=-q\ddot{q}$ which holds on the light cone. We see that the Hessian
has two independent eigenvectors $q$ and $w$ both to the eigenvalue $0$. The
Hessian rank is thus $4-2=2$ as it should be for a particle constrained to a
sphere. The Hessian constraints are $\dot{q}q=0$ and $qq=0$. They commute with
taking the projection. The corresponding primary constraints are $pq=0$ and $qq=0$.
These two constraints are the primary constraints used to construct the Hamiltonian
in the Dirac formalism. Alternatively, we could have introduced the internal
coordinates on a sphere and find the ordinary Hamiltonian (in that case the
corresponding Hessian would be a nonsingular $2\times2$ matrix, that is, with rank
$2$).
\end{small}

\end{document}